\documentclass[a4paper, 10pt, onecolumn]{article}      

\usepackage{amsfonts}
\usepackage{amssymb}
\usepackage[dvips]{graphicx}
\usepackage{epsfig}
\newtheorem{remark}{Remark}
\newtheorem{proposition}{Proposition}
\newtheorem{proof}{Proof}

\title{\LARGE \bf
On Designing Lyapunov-Krasovskii Based AQM for Routers
Supporting TCP Flows}

\author{Yann Labit$^{\ast\dag}$, Yassine Ariba$^{\ast\dag}$ and Fr\'ed\'eric Gouaisbaut$^{\ast}$
\thanks{ Universit\'e de Toulouse; UPS, 118 Route de Narbonne, F-31062 Toulouse, France.}%
\thanks{ LAAS; CNRS;  7, avenue du Colonel Roche, F-31077 Toulouse, France.
 {\tt\small ylabit@laas.fr, yariba@laas.fr, fgouaisb@laas.fr}}%
}

\date{December 2007}

\begin{document}

\maketitle
\thispagestyle{empty}
\pagestyle{empty}

\begin{abstract}
For the last few years, we assist to a growing interest of designing
AQM (Active Queue Management) using control theory. In this paper,
we focus on the synthesis of an AQM based on the Lyapunov
theory for time delay systems. With the help of a recently developed
Lyapunov-Krasovskii functional and using a state space
representation of a linearized fluid model of TCP, two
 robust AQMs stabilizing the TCP model are constructed. Notice that our results are constructive and the synthesis
problem is reduced to a convex optimization scheme expressed in
terms of linear matrix inequalities (LMIs). Finally, an example
extracted from the literature and simulations via {\it NS simulator} \cite{Fal}
support our study.
\end{abstract}

\section{Introduction}
Over a past few years, problems have arisen with regard to Quality
of Service (QoS) issues in Internet traffic congestion control
\cite{Low02}, \cite{Sri04}. AQM mechanism, which
supports the end-to-end congestion control mechanism of Transmission
Control Protocol (TCP), has been actively studied by many
researchers. AQM controls the queue length of a router by
actively dropping packets. Various
mechanisms have been proposed in the literature such as
Random Early Detection (RED) \cite{Flo93}, Random Early Marking
(REM) \cite{Ath00},
 Adaptive Virtual Queue (AVQ) \cite{Kun01}
and many others \cite{Ryu04}. Their performances have been evaluated \cite{Fir00}, \cite{Ryu04}
and empirical studies have shown the effectiveness of these algorithms \cite{Le03}. Then, significant research
has been devoted to the use of control theory to develop more
efficient AQMs. Using dynamical model developed by \cite{Mis00},
some P ({\it Proportional}), PI ({\it Proportional Integral})
\cite{Hol02} have been designed as well as robust
control framework issued \cite{Que04}. Nevertheless, most
of these papers do not take into account the delay and ensure the
stability in closed loop for all delays which could be
conservative in practice.\\~\indent
The study of congestion problem with time delay systems framework is not new and has been succesfully
exploited. In \cite{Mic06}, \cite{Pap04}, using Lyapunov-Krasovskii theory, the global stability analysis of
the non linear model of TCP is performed. In \cite{Kim06}, a delay dependent state feedback controller is provided by
compensation of the delay with a memory feedback control. This latter methodology is interesting in theory but hardly suitable
in practice.\\~\indent
 Based on a recently developed Lyapunov
functional for time delay systems, two AQMs stabilizing
the TCP model are constructed. The first one is called IOD-AQM ({\it Independent Of
Delay}) and it deals with the robust control of TCP for all delays
in the loop. The second one, DD-AQM ({\it Delay Dependent}) is
devoted to the control of the TCP dynamics when an upperbound of
the delay is known. In order to consider a more realistic case,
extension to the robust case, where the delay is uncertain is considered using quadratic
stabilization framework.\\~\indent The paper is organized as
follows. The second part presents the uncertain mathematical model
of a network supporting TCP. Section III is dedicated to the
design of two AQMs ensuring the robust stabilization of TCP. Section IV presents application
of the exposed theory and the simulation results
using NS-2. Finally, section V concludes the paper.
\\~\indent {\it Notations:} For
two symmetric matrices, $A$ and $B$, $A>$ ($\geq$) $B$ means that
$A-B$ is (semi-) positive definite. $A^T$ denotes the transpose of
$A$. $\sf{1}_n$ and $\sf{0}_{m\times n}$ denote respectively the
identity matrix of size $n$ and null matrix of size $m\times n$. If
the context allows it, the dimensions of these matrices are often
omitted. For a given matrix $A\in\sf{R}^{n\times n}$, $<A>$ stands for $A+A^T$.

\section{Problem statement}
\subsection{The linearized fluid-flow model of TCP}
The fluid flow model of TCP considered here was
introduced in \cite{Mis00}, \cite{Hol02}. Based on this system, we
will construct two AQM, which take into account
delays inherent to networks.\\~\indent Given the network
parameters: number of TCP sessions, link capacity and propagation
delay ($N$, $C$ and $T_p$ respectively), we define the set of
operating points $(W_0,q_0,p_0)$ by $\dot{W}=0$ and $\dot{q}=0$:
\begin{equation}
  \label{point_eq}
    \left\{\begin{array}{l}
\displaystyle\dot{W}=0~\Rightarrow~W_0^2p_0=2\\
\displaystyle\dot{q}=0~\Rightarrow~W_0=\frac{R_0C}{N},~R_0=\frac{q_0}{C}+T_p
\end{array}\right.
\end{equation}
~\indent where $W(t)$ is the congestion window, $q(t)$ is the queue length at the congested router and $R(t)$ is the Round Trip Time (RTT) which represents the delay in TCP dynamics. $x_0$ denotes the value of the variable $x$ at the equilibrium point.\\
~\indent Assuming $N(t)\equiv N$ and $R(t)\equiv R_0$ as constants,
the dynamic model of TCP can be approximated, around an
equilibrium point, by the linear time delay system \cite{Mis00}:
\begin{equation}
  \label{sys_lin}
  \left\{\begin{array}{ll}
\displaystyle \delta\dot{W}(t)=-\frac{N}{R_0^2C}\Big(\delta W(t)+\delta W(t-h(t))\Big)\\ \hspace*{0.7cm}-\frac{1}{R_0^2C}\Big(\delta q(t)-\delta q(t-h(t))\Big)-\frac{R_0C^2}{2N^2}\delta p(t-h(t))\\
\displaystyle\delta\dot{q}(t)=\frac{N}{R_0}\delta
W(t)-\frac{1}{R_0}\delta q(t)
\end{array}\right.
\end{equation}
where $\delta W \doteq W-W_0$, $\delta q \doteq q-q_0$ and $\delta p
\doteq p-p_0$ are the state variables and input perturbations
around the operating point. The model (\ref{sys_lin}) is valid only
if the variations of these new variables are kept enough
small.\\~\indent
The input of our model (\ref{sys_lin}) corresponds to the drop probability of a packet.
 This probability is fixed by the AQM. This latter has for objective to regulate the queue size at the router.\\~\indent
For synthesis problem (see section \ref{preliminaries}), we
consider a state feedback. So that, the queue
management strategy of the drop probability will be expressed as
\begin{equation}
p(t)=p_0+k_1\delta W(t)+k_2\delta q(t).
\end{equation}

\begin{remark}
~\\
~\indent i) It is possible to design a state feedback as it corresponds to a PD ({\it Proportional Derivative}) control law \cite{Kim06}. Furthermore, although $W(t)$ is not measured, one can estimate $y(t) =
\frac{W(t)}{R_0}$, the aggregate flow at the link \cite{Kim06}, \cite{Low02}.\\
~\indent ii) The main difficulty in all representations of TCP behavior is
the exact estimation of network parameters (and not the state
feedback control law). Two techniques
are used:\\
$\bullet$ Active measurements \cite{Lab05r}, \cite{Pra03} consist in generating probe traffic in the network, and
then observing the impact of network components and protocols on
traffic: loss rate, delays, RTT, capacity... Therefore, as active
measurement tools generate traffic in the network (intrusiveness), one of their
major drawbacks is related to the disturbance introduced by the
probe traffic which can make the network QoS change, and thus
provide erroneous measures \cite{Lab05r}. Sometimes, active probing
traffic can be seen as denial of service attacks (DoS), scanning, or
something else but in any case as hacker acts. Probe traffic is then
discarded, and its source can be blacklisted.\\
 $\bullet$ Passive measurements refer to the
process of measuring a network, without generating or modifying any
traffic on the network. Passive monitoring is done with the capture
of traffic and estimate off line networks parameters: It's still to
be non intrusive (good estimation of parameters) but not reactive.
The passive evaluation relies on DAG system cards \cite{Cle00} that
represent references for such kind of measurements.

Passive and active measurements is still a growing interest because
exact estimation of networks parameters still difficult since the
heterogeneity of autonomous systems \cite{Lab05r}. A future idea
(early introduced in this paper) for this problem is to consider
uncertainties for parameters: This solution allows to use robust
control theory in sense of polytops.
\end{remark} 

\subsection{Time delay system approach}
In this paper, we choose to model the dynamics of the queue and the
congestion window as a time-delay system. Indeed, the delay is an
intrinsic phenomenon in networks. Taking into account this
characteristic, we expect to reflect as much as possible the TCP
behavior, providing more relevant analysis and synthesis
methods.\\~\indent The linearized TCP fluid model (\ref{sys_lin})
can be rewritten as the following time delay system:
\begin{equation}
\label{forme_can} \left\{\begin{array}{l}
\dot{x}(t)=Ax(t)+A_dx(t-h)+Bu(t-h)\\
x_0(\theta)=\phi(\theta),~\mbox{with }\theta\in[-h,0]\end{array}\right.
\end{equation}
with
\begin{equation}
\label{esp_etat}
{\small A\!=\!\left[\begin{array}{cc}\! -\frac{N}{R_{0}^{2}C}\! &\!
-\frac{1}{CR_{0}^2}\! \\\! \frac{N}{R_0}\! &\! -\frac{1}{R_0}\!
\end{array}\right]\!,\!~A_d\!=\!\left[\begin{array}{cc}\! -\frac{N}{R_{0}^{2}C}\! &\!
\frac{1}{R_{0}^2C}\! \\\! 0\! &\! 0\! \end{array}\right]\!,\!B\!=\!\left[\begin{array}{c}\!
-\frac{C^{2}R_0}{2N^2}\!\\\! 0\! \end{array}\right]\!.}
\end{equation}
where $x(t)=[\delta W(t) ~~\delta
q^T(t)]$ is the state vector and $u(t)=\delta p(t)$ the input.
$\phi(\theta)$ is the initial condition.\\~\indent There are mainly
three methods to study time delay system stability: analysis of the
characteristic roots, robust approach and Lyapunov theory. The
latter will be considered because it is an effective and practical
method which provides LMI (Linear Matrix Inequalities
\cite{Boy94}) criteria. To analyze and control the system
(\ref{forme_can}), the Lyapunov-Krasovskii approach \cite{Gu03} is
used which is an extension of the traditional Lyapunov
theory.\\~\indent In the literature, few articles using time delay
systems approach to model TCP dynamic already appeared. In
\cite{Wan03}, a delay dependent robust stability condition was
proposed and the design of a state feedback was derived. However,
the criterion used is quite obsolete and thus conservative. Then,
other papers design control laws based on predictor \cite{Kim06}.
The predictive approach is an interesting method theoretically but
not in practice, moreover the delay has to be known exactly.
\cite{Mic06} and \cite{Pap04} use time
delay system approach too and propose global stability analysis of the linear model.
However synthesis is not considered.\\
~\indent In this paper, we aim at providing
 methods to control system
(\ref{forme_can}) with different objectives: giving conditions for
the nominal or robust stabilization for {\it IOD} and {\it DD} cases.

\subsection{Polytopic uncertain model}
\label{pol_section}
The state space representation shows that matrices $A$, $A_d$
and $B$ depend on network parameters. Especially, it depends on the
RTT $R_0$, a significant parameter, which is difficult to
estimate in practice. For a more rigorous study, it could be
interesting to take into account some uncertainty on the delay
$R_0$.\\~\indent Let then rewrite system (\ref{forme_can}) as
following
\begin{equation}
\label{sys_inc} \dot{x}(t)=A(R_0)x(t)+A_d(R_0)x(t-h)+B(R_0)u(t-h).
\end{equation}
~\indent With the polytopic approach, the idea is to insure the
stability for a set of systems. Let suppose that
$R_0\in[R_{0_{min}},R_{0_{max}}]$, then matrices $A$, $A_d$ and
$B$ belong to a certain set
$$ \Omega=\{[A,A_d,B]~|~R_0\in[R_{0_{min}},R_{0_{max}}]\}$$
and we aim at looking for an AQM (expressed in term of state
feedback) which stabilizes system (\ref{sys_inc}) for all matrices
belonging to $\Omega$. However, the parameter $R_0$ doesn't appear
linearly in the matrices $A$, $A_d$ and $B$. So that, the set
$\Omega$ defined by the uncertainty is non
convex. \\~\indent A common idea in robust control theory is to look for
a polytopic set $\mathcal{P}$ which includes the set $\Omega$. Using
convexity property, it is much more easy to test the stability in
closed loop for the overall polytop. If the stability of
$\mathcal{P}$ is proved, then the stability of $\Omega$ is
insured.\\
~\indent In order to create the polytop $\mathcal{P}$, we pose
$\rho_{1}=\frac{1}{R_0}$, $\rho_{2}=\frac{1}{R_0^2}$ and
$\rho_{3}=R_0$. Since there are three uncertain parameters, the
polytop will have $n_\omega=8$ vertices. For a bounded value $R_0$,
the new uncertain parameters $\rho_i$, $\forall i=\{1,2,3\}$ are
bounded. So, the matrices of the uncertain system (\ref{sys_inc})
are defined as
\begin{equation}
\label{def_mat_pol}
\begin{array}{l}
A=\rho_1 \left[\begin{array}{cc} 0&0\\N&-1 \end{array}\right] + \rho_2\left[\begin{array}{cc} -\frac{N}{C}&-\frac{1}{C}\\0&0 \end{array}\right]=\rho_1 A_0+\rho_2 A_1,\\
A_d=\rho_2 \left[\begin{array}{cc} -\frac{N}{C}&\frac{1}{C}\\0&0 \end{array}\right]=\rho_2 A_{d_0},\\
B=\rho_3 \left[\begin{array}{c} -\frac{C^2}{2N}\\0\end{array}\right]=\rho_3 B_0.
\end{array}
\end{equation}
~\indent The set $\Omega$ is contained in the set $\mathcal{P}$,
$$ \Omega \subset co\{\omega^{(i)},~i=1,2,...,8\}.$$
~\indent where the $\omega^{(i)}$ are the vertices of $\mathcal{P}$.

\section{Stabilization using time-delay system approach}
\label{preliminaries}

In the previous section, an uncertain model of the TCP/AQM dynamic has been designed. This section is devoted to the construction of robust
AQM stabilizing a such model. The first approach proposes the
construction of an independent of delay ({\it IOD}) controller using convex optimisation schemes (LMI). In a second part, we
describe a delay dependent ({\it DD}) method which takes into account
the size of the delay. Using an information on the delay, we expect a reduction of conservatism and then an improvment of
results.

\subsection{{\it Independent of delay} AQM design}

The idea is to insure the stability in closed loop for all delays
as it has been proposed in \cite{Hol02} using frequential
arguments and traditionnal control tools. Here, we propose to use the following well-known
Lyapunov-Krasovskii functionnal:
\begin{equation}
\label{eq_iod2}
V(x_t)=x^T(t)Px(t)+\int_{t-h}^{t}x^T(\theta)Qx(\theta)d\theta.
\end{equation}
where the matrices $P$ and $Q$ are symmetric and positive definite.
The choice of this Lyapunov-Krasovskii functional implies the
following proposition. \begin{proposition} \cite{Gu03} \label{prop_ana_iod}
System (\ref{forme_can}) for $u(t)=0$ is asymptotically stable
$\forall h\geq 0$ if there exist real symmetric matrices $P>0$ and
$Q$, such that
\begin{equation}
\label{lmi_iod} \left[\begin{array}{cc} A^T{\bf P}+{\bf P}A+{\bf Q} & {\bf P}A_d
\\ A_d^T{\bf P} & -{\bf Q} \end{array}\right] < \sf{0}
\end{equation}
is satisfied. \end{proposition} ~\indent Now, we construct the following
memoryless state feedback
\begin{equation}
\label{ret_etat} u(t)=Kx(t),~K\in\sf{R}^{m\times n}
\end{equation}
~\indent to control system (\ref{forme_can}) ($K$ is a constant matrix gain). This controller corresponds to our AQM.\\
~\indent Applying (\ref{ret_etat}) to (\ref{forme_can}), we get the
closed-loop system
\begin{equation}
\label{closed_loop} \left\{\begin{array}{ll}
\dot{x}(t)=Ax(t)+\widetilde{A_d}x(t-h)\\
x_0(\theta)=\phi(\theta),~\mbox{with}~\theta\in[-h,0]
\end{array} \right.
\end{equation}
with $\widetilde{A_d}=A_d+BK$.\\~\indent
Then, the following synthesis criterion can be easily derived from (\ref{lmi_iod}) and (\ref{closed_loop}).
\begin{proposition} \label{prop_syn_iod}
If there exist symmetric matrices $R>0$, $S$ and a matrix
$Z\in\sf{R}^{m\times n}$, such that
\begin{equation}
\label{mat_iod} \left[\begin{array}{cc} {\bf R}A^T+A{\bf R}+{\bf S} & A_d{\bf
R}+B{\bf Z} \\ {\bf R}A_d^T+{\bf Z^T}B^T & -{\bf S} \end{array}\right]  < \sf{0}
\end{equation}
then, system (\ref{forme_can}) is stable under the control law
(\ref{ret_etat}) with the feedback gain $K=ZR^{-1}$. \end{proposition}
~\indent This latter proposition provides an IOD-AQM, $K$, which stabilize (\ref{forme_can}) for all delays $h\in\sf{R}^+$.
\subsection{{\it Delay dependent} AQM design}

In this subsection, our goal is to design a controller which takes into account the upperbound of the delay. The delay dependent case starts from a system stable without delays and looks for the maximal delay that preserves stability.\\
~\indent Generally, all methods involve a Lyapunov functional,
and more or less tight techniques to bound some cross terms and to
transform system \cite{Gu03}. These choices of specific
Lyapunov functionals and overbounding techniques are the origin of
conservatism. In the present paper, we choose a recent
Lyapunov-Krasovskii functional (\ref{lyap_ddr})
\cite{Gou06b}:
\begin{equation}
\label{lyap_ddr} \begin{array}{c}
\displaystyle V(x_t)=x^T(t)Px(t)+\int\limits_{t-\frac{h}{r}}^{t}\!\!\int\limits_{\theta}^{t}\dot{x}^T(s)R\dot{x}(s)dsd\theta\\[0.5em] \displaystyle~~~~~+\int_{t-\frac{h}{r}}^{t}\left(\begin{array}{c} x(s)\\x(s-\frac{1}{r}h)\\\vdots\\x(s-\frac{r-1}{r}h)\end{array}\right)^TQ\left(\begin{array}{c} x(s)\\x(s-\frac{1}{r}h)\\\vdots\\x(s-\frac{r-1}{r}h)\end{array}\right)ds
\end{array}
\end{equation}
~\indent where $P\in\sf{S}^n$ is a positive definite matrix,
$Q\in\sf{S}^{rn}$ and $R\in\sf{S}^n$ are two positive definite
matrices. $r\geq 1$ is an integer corresponding to the discretization
step. Using this functional, we propose the following. \begin{proposition}
\label{prop_dd} If there exist symmetric positive definite matrices
$P$, $R\in\sf{R}^{n\times n}$, $Q\in\sf{R}^{rn\times rn}$, a matrix
$X\in\sf{R}^{(r+2)n\times n}$, a scalar $h_{m}>0$, an integer $r\geq 1$
and a matrix $K\in\sf{R}^{m\times n}$ such that
\begin{equation}
\label{bmi_dd}  \Gamma+{\bf X}S+S^T{\bf X}^T< \sf{0}
\end{equation}
~\indent where
\begin{equation}
\label{gamma} {\small \Gamma=\!\left[\begin{array}{ccccc}\!\! \frac{h_{m}}{r}{\bf
R}\!\!&\!\!{\bf
P}\!\!&\!\!\sf{0}\!\!&\!\!\ldots\!\!&\!\!\sf{0}\!\!\\\!\!{\bf
P}\!\!&\!\!-\frac{r}{h_{m}}{\bf R}\!\!&\!\!\frac{r}{h_{m}}{\bf
R}\!\!&\!\!~\!\!&\!\!\vdots\\\!\!\sf{0}\!\!&\!\!\frac{r}{h_{m}}{\bf
R}\!\!&\!\!-\frac{r}{h_{m}}{\bf
R}\!\!&\!\!~\!\!&\!\!\vdots\!\!\\\!\!\vdots\!\!&\!\!~\!\!&\!\!~\!\!&\!\!\ddots\!\!&\!\!\vdots\!\!\\\!\!\sf{0}\!\!&\!\!\ldots\!\!&\!\!\ldots\!\!&\!\!\ldots\!\!&\!\!\sf{0}\!\!
\end{array}\right]\!\!+\!\!\left[\begin{array}{ccc}\!\!
\sf{0}\!\!&\!\!\ldots\!\!&\!\!\sf{0}\!\!\\\!\!\vdots\!\!&\!\!{\bf
Q}\!\!&\!\!\vdots\!\!\\\!\!\sf{0}\!\!&\!\!\ldots\!\!&\!\!\sf{0}\!\!
\end{array}\right]\!\!+\!\!\left[\begin{array}{cc}\!\!
\sf{0}\!\!&\!\!\ldots\!\!\\\!\!\sf{0}\!\!&\!\!\ldots\!\!\\\!\!\vdots\!\!&\!\!{\bf
Q}\!\! \end{array}\right]}
\end{equation}
~\indent and $S=\left[\begin{array}{cccc} -\sf{1} & A &\sf{0}_{n\times (r-1)n}&\widetilde{A_d} \end{array}\right]$\\
then, system (\ref{forme_can}) can be stabilized for all $h\leq
h_{m}$ by the control law $u(t)=Kx(t)$. \end{proposition}

\begin{proof} It is always possible to rewrite
(\ref{closed_loop}) as $S\xi=\sf{0}$ where
 \begin{equation}
   \label{proj_ddr}
\begin{array}{c}
   \xi= \left[\begin{array}{c} \dot{x}(t)\\x(t)\\x(t-\frac{1}{r}h)\\ \vdots\\x(t-\frac{r-1}{r}h)\\x(t-h) \end{array}\right]\in\sf{R}^{(r+2)n}\\~\mbox{and}~S=\left[\begin{array}{cccc} -\sf{1} & A &\sf{0}_{n\times (r-1)n}&\widetilde{A_d} \end{array}\right]
\end{array}
 \end{equation}
~\indent Using the extended variable $\xi(t)$ (\ref{proj_ddr}), the
derivative of $V$ along the trajectories of system (\ref{forme_can})
leads to:
 \begin{equation}
 \left\{\begin{array}{l}
   \dot{V}(x_t)=\xi^T \left[\begin{array}{ccccc} \frac{h}{r}{\bf R}&{\bf P}&\sf{0}&\ldots&\sf{0}\\{\bf P}&-\frac{r}{h}{\bf R}&\frac{r}{h}{\bf R}&~&\vdots\\\sf{0}&\frac{r}{h}{\bf R}&-\frac{r}{h}{\bf R}&~&\vdots\\\vdots&~&~&\ddots&\vdots\\\sf{0}&\ldots&\ldots&\ldots&\sf{0} \end{array}\right]\xi\\ \hspace*{0.7cm}+\xi^T \left[\begin{array}{ccc} \sf{0}&\ldots&\sf{0}\\\vdots&{\bf Q}&\vdots\\\sf{0}&\ldots&\sf{0} \end{array}\right]\xi-\xi^T \left[\begin{array}{cc} \sf{0}&\ldots\\\sf{0}&\ldots\\\vdots&{\bf Q} \end{array}\right]\xi < \sf{0}\\
 \mbox{such that} \left[\begin{array}{cccccc} -\sf{1}&A&\sf{0}&\cdots&\sf{0}&\widetilde{A_d} \end{array}\right] \xi=0
 \end{array} \right.
 \end{equation}
\begin{equation}
   \label{proj_ddr2}
\Leftrightarrow \left\{\begin{array}{ll}
  \dot{V}(x_t)=\xi^T\Gamma\xi< \sf{0}\\
 \mbox{such that} \left[\begin{array}{cccccc} -\sf{1}&A&\sf{0}&\cdots&\sf{0}&\widetilde{A_d} \end{array}\right] \xi=0
 \end{array} \right.
\end{equation}
~\indent where $\Gamma\in\sf{S}^{(r+2)n}$ depends on $P$, $R$, $Q$
and the delay $h$.\\~\indent
Using projection lemma \cite{Ske98}, there exists $X\in\sf{R}^{(r+2)n\times n}$ such that (\ref{proj_ddr2}) is equivalent to (\ref{bmi_dd}).\end{proof}
\begin{remark}
~\\
\begin{itemize}
\item There exists another equivalent form of this LMI in term of analysis (i.e. with $K=\sf{0}$) provided in \cite{Gou06b} and based on robust control tools.
\item In term of analysis, it is shown in \cite{Gou06a} that for $r=1$, this proposed function (\ref{lyap_ddr}) is equivalent to the main classical results of the literature. Moreover, in the same study it is proved that for $r>1$ conservatism is reduced.
\end{itemize}
\end{remark}
~\indent Nevertheless, applying a state feedback (\ref{ret_etat}),
we have $\widetilde{A_d}=A_d+BK$ with the controller gain $K$ appearing
as a decision variable. Then, the condition becomes a BMI. That's the reason why in this paper, we propose a relaxation algorithm. The algorithm principle
consists to alternate analysis and synthesis steps.\\~\indent First
let define the {\it synthesis LMI}:
 \begin{equation}
   \label{lmi_syn}
{\small  \Gamma\!+<X
\left[\begin{array}{cccccc}\!-\sf{1}\!&\!A\!&\!\sf{0}\!&\!\cdots\!&\!\sf{0}\!&\!A_d\!+\!B{\bf K}\!
\end{array}\right]><\! \sf{0}}
 \end{equation}
~\indent where $K$ $\in\sf{R}^{m\times n}$ and
$X$ is the slack variable which has been
fixed.\\~\indent
By the same way, we define the {\it analysis LMI}:\\
 \begin{equation}
   \label{lmi_ana}
{\small
   \Gamma+<{\bf X} \left[\begin{array}{cccccc}-\sf{1}&A&\sf{0}&\cdots&\sf{0}&A_d+BK \end{array}\right]>< \sf{0}}
 \end{equation}
~\indent where $K$ is fixed. Then, we propose the following
algorithm.\\~\indent \underline{{\bf Algorithm}}:\\~\indent
$\bullet~\mbox{Slack variable
initialization,}~X=X_0$
\begin{enumerate}
\item  We solve the {\it synthesis} optimization
$$\left\{\begin{array}{l}
\displaystyle h_{max_i}^s=\max_{{\bf P,Q,R,K_i}}\{h_{m}\}\\
\mbox{s.t.}~LMI~(\ref{lmi_syn})
\end{array} \right.$$
~\indent A matrix gain called $K_i$ is derived.
\item  We solve the {\it analysis} optimization with $K=K_i$.
$$\left\{\begin{array}{l}
\displaystyle h_{max_i}^a=\max_{{\bf P,Q,R,X_i}}\{h_{m}\}\\
\mbox{s.t.}~LMI~(\ref{lmi_ana})
\end{array} \right.$$
~\indent The new slack variable is derived $X_i$.
\end{enumerate}
~\indent$\bullet$ We test if $h_{{max}_i}^a=h_{{max}_i}^s$.
\begin{itemize}
\item if true, there is no improvement on the maximal size of the allowable delay: end of the algorithm.
\item if false, the process is reiterated to the step (1) with a new slack variable and upperbound of the delay.
\end{itemize}
\begin{remark} At the {\it test} step, one always has $h_{{max}_i}^a\geq h_{{max}_i}^s$. Consequently, throughout the progression of the algorithm the upperbound $h_m$ can not regress.\end{remark}
~\indent Notes that the main problem, which is common in
relaxation methods, remains the initialization of slack variables.

\section{Application to TCP/AQM dynamics and validation through NS-2}


In this section, we are going first to consider the nominal system
in order to expose the control principle. Then, we will
extend our methods to the robust case. For a realistic
case, it is essential to insure stability in spite of the delay
uncertainty.

\subsection{Numerical example}

As a widely adopted numerical illustration extracted from
\cite{Hol02}, consider the case when $q_0=175$ packets, $T_p=0.2$
 second and $C=3750$ packets/s (corresponds to a $15$ Mb/s link with average packet size $500$ bytes). Then, for a load of $N=60$ TCP sessions, we have $W_0=15$ packets, $p_0=0.008$, $R_0=0.246$ seconds. We obtain the following open loop system
\begin{equation}
\label{ex_num1} \begin{array}{l}
\left[\begin{array}{c}\delta\dot{W}(t)\\\delta\dot{q}(t)\end{array}\right]=\left[\begin{array}{cc} -0.2644 &
-0.0044 \\ 243.9024 & -4.0650 \end{array}\right] \left[\begin{array}{c}\delta W(t)\\\delta
q(t)\end{array}\right] \\ \hspace*{0.8cm}+\left[\begin{array}{cc} -0.2644 & 0.0044 \\ 0 & 0
\end{array}\right]\left[\begin{array}{c}\delta W(t-h(t))\\\delta q(t-h(t))\end{array}\right]\end{array}
\end{equation}
Matrix $A$ is Hurwitz and applying {\it IOD} proposition
\ref{prop_ana_iod}, we observe that the LMI (\ref{lmi_iod}) is
feasible. So, we conclude (\ref{ex_num1}) is {\it IOD} stable and
system (\ref{ex_num1}) is stable for all $h\in\sf{R}^+$. However, in order to avoid congestion and to regulate the queue size at a desired level in spite of uncertainty on delay,
an AQM has to be implanted.

\subsection{{\it IOD}/{\it DD} Synthesis}

\subsubsection{Independent of delay method}

In the {\it IOD} case, for nominal system only, it turns out that
the delayed term, which can be viewed as a disturbance, can be eliminated choosing $k_1$ and $k_2$ as:
\begin{equation}
\label{gain_k} K=[k_1~~k_2]~\mbox{such that}~\left\{\begin{array}{l}
\displaystyle k_1=-\frac{2N^3}{R_0^3C^3}\\
\displaystyle k_2=\frac{2N^2}{R_0^3C^3}
\end{array} \right.
\end{equation}
~\indent  Thus if $A$ is Hurwitz and
for a state feedback gain $K$ defined as (\ref{gain_k}), then the
system (\ref{forme_can}) is {\it IOD} stable. Since an {\it IOD} stabilizing gain K can always be found (in nominal case), this method provides a systematic technique for the algorithm initialization (for {\it DD} synthesis).\\~\indent Concerning the
robustness issue, let consider that $R_0$ is uncertain such
that $R_{0_{min}}\leq R_0\leq R_{0_{max}}$. This system will be
stable if the polytop $\mathcal{P}$ (see section \ref{pol_section}) is
stabilized. Using the quadratic stability framework \cite{Boy94}, we propose the
following result. \begin{proposition} \label{prop_syn_iod_pol} The system
(\ref{sys_inc}) will be quadratically stabilized with the state
feedback $u(t)=Kx(t)$ $\forall h\geq 0$ and $\forall
R_0\in[R_{0_{min}},R_{0_{max}}]$, if there exist symmetric matrices
$R>0$, $S$ and a matrix $Z\in\sf{R}^{m\times n}$, such that
\begin{equation}
\label{mat_iod_pol} \left[\begin{array}{cc} {\bf R}A^{{(i)}^T}+A^{(i)}{\bf R}+{\bf
S} & A_d^{(i)}{\bf R}+B^{(i)}{\bf Z} \\ {\bf R}A_d^{{(i)}^T}+{\bf
Z^T}B^{{(i)}^T} & -{\bf S} \end{array}\right]  < \sf{0}
\end{equation}
is satisfied for all $i=\{1,2,...,8\}$. The matrix gain is given by
$K=ZR^{-1}$.  $A^{(i)}$, $A_d^{(i)}$ and $B^{(i)}$ correspond
to the matrices values on the vertices $\omega^{(i)}$ (\ref{def_mat_pol}):
\begin{equation}\label{sommets} \omega^{(i)}=(A^{(i)},A_d^{(i)},B^{(i)}), i=1,2,...,8\end{equation}\end{proposition}
~\indent Taking numerical example (\ref{ex_num1}), $R_0$ nominal
value is $0.246$ seconds. The objective is to stabilize the uncertain system (\ref{sys_inc})
for given bounds ($R_{0_{min}}$ and $R_{0_{max}}$), and to
maximize the stability domain. \\~\indent
Applying {\it IOD} LMI condition of the proposition
\ref{prop_syn_iod_pol}, we obtain the results of table I.
\begin{table}[!h]
\label{iod_pol} \centering
\begin{tabular}{|c|c|c|}
\hline
 ${\bf R_{0_{min}}}$ & ${\bf R_{0_{max}}}$ &{\bf Gain} ${\bf K}$ \\
\hline $0.1$ & $0.4$ &
$10^{-3}[-0.3709~~0.0062]$ \\
\hline
$0.15$ & $0.83$ & $10^{-4}[-0.4729~~0.0079]$ \\
\hline
\end{tabular}\\
\caption{{\it IOD} state feedback gains to stabilize a polytop}
~\vspace*{-0.9cm}
\end{table}
\\~\indent We can observe, in that case, decreasing the lowerbound is more restrictive than increasing the upperbound for the optimization problem. Nevertheless, because of the unavoidable delay in networks (like propagation delays), it is useless to look for a very small lowerbound.
\subsubsection{ Delay-dependent method}
\label{algo_stab} Using the relaxation algorithm previously exposed and {\it IOD} gain (\ref{gain_k}) for the initialization,
we get the following results of the table II for the robust delay
dependent case where a common Lyapunov-Krasovskii functional is
found for each vertice of the polytop. Compared to {\it IOD}
results, we improve sligthly the set of admissible delays.
\begin{table}[!h]
\label{dd_pol} \centering
\begin{tabular}{|c|c|c|c|c|}
\hline
 ${\bf r}$ & ${\bf [R_{0_{min}},R_{0_{max}}]}$ &{\bf Gain} ${\bf K}$ & ${\bf h_{m}}$ \\
\hline $1$ & $[0.1,0.45]$ & $10^{-3}[-0.589~~0.0244]$ &
$0.56$ \\
\hline $1$ & $[0.1,0.5]$ & $10^{-3}[-0.321~~0.0204]$ &
$0.48$ \\
\hline
$2$ & $[0.1,0.45]$ & $10^{-3}[-0.575~~0.0240]$ & $0.62$ \\
\hline
$2$ & $[0.1,0.5]$ & $10^{-3}[-0.272~~0.0193]$ & $0.52$ \\
\hline
\end{tabular}\\
\caption{{\it DD} state feedback gains to stabilize a polytop}
~\vspace*{-0.9cm}
\end{table}
\begin{remark}
~\\
\begin{itemize}
\item If $R_{0_{max}}>h_m$, then system (\ref{sys_inc}) is just stable for $R_0\in[R_{0_{min}},h_m]$ since $R_0$ is the RTT and corresponds to the delay.
\item As expected, we obtain better results for $r=2$, since $h_{max}$ is larger.
\end{itemize}
\end{remark}
Our results can be compared with results from \cite{Wan03} where a robust delay dependent stabilization is designed. In \cite{Wan03}, the system in closed loop is shown to be robustly stable for $R_0\in[0,0.216]$ while the proposed criterion of proposition \ref{prop_dd} robustly stabilises the system for $R_0\in[0.1,0.5]$.

\subsection{Simulations}

\begin{figure}[!h]
\begin{center}
  \includegraphics[angle=0,width= 6.5cm]{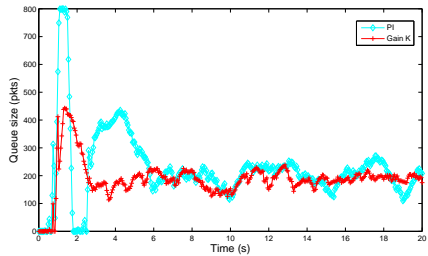}
\caption{\it Time evolution of the queue length: comparison between
PI and state feedback Gain K.} \label{KPI}
\end{center}
\end{figure}

\begin{figure}[!h]
\begin{center}
  \includegraphics[angle=0,width= 6.5cm]{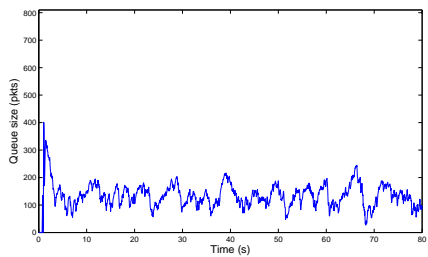}
\caption{\it Time evolution of the queue length for gain $K$
(calculated from {\it DD} robust stabilization)with a perturbation on the delay.} \label{dd_a1_per}
\end{center}
\end{figure}

\begin{figure}[!h]
\begin{center}
  \includegraphics[angle=0,width= 6.5cm]{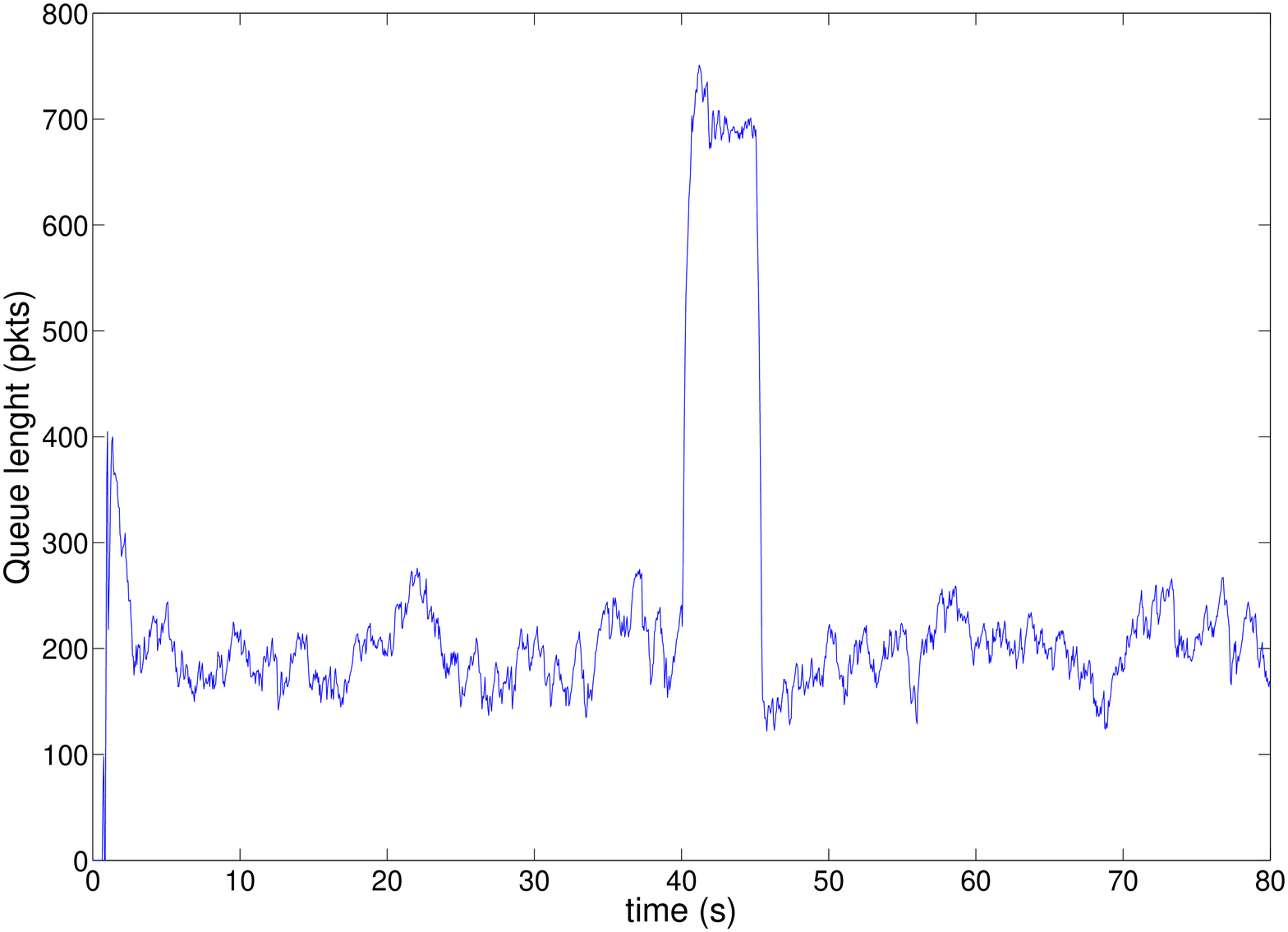}
\caption{\it Time evolution of the queue length for gain $K$
(calculated from {\it DD} robust stabilization) with a perturbation as UDP traffic.}
\label{dd_a2_per}
\end{center}
\end{figure}

We aim at proving the effectiveness of our method using NS-2
\cite{Fal}, a network simulator widely used in the communication
community. Taking values from the previous numerical example, we
apply the new AQM based on a state feedback (i.e a simple constant
matrix gain $K$). The target queue length $q_0$ is $175$ packets
while buffer size is $800$. The average packet length is $500$
bytes. The default transport protocol is TCP-New Reno without ECN
marking.\\~\indent For the convenience of comparison, we adopt the
same values and network configuration than \cite{Hol02} who design
a PI controller ({\it Proportional-Integral}). This PI is configured
as follow, the coefficients $a$ and $b$ are fixed at $1.822e-5$ and
$1.816e-5$ respectively, the sampling frequency is
$160$Hz.\\~\indent In the figure \ref{KPI}, we apply the gain $K$
from the table II which ensures {\it DD} robust stability. We compare our result with PI AQM
provided by \cite{Hol02}. It appears that our control allows a faster response as
well as a smaller overshoot.
~\indent Simulations of perturbed system is reported in figures \ref{dd_a1_per} and \ref{dd_a2_per}. In figure \ref{dd_a1_per}, we have increased the propagation delay by $20$ ms. Even if the system converges to a different reference point (slightly lower), the queue size is stable and quickly regulated.
~\\~\indent For more important pertubations (on the delay $R_0$ or number of
sessions $N$), the system in closed-loop is still stable but the steady state changes since we converge to a new equilibrium point. In figure \ref{dd_a2_per}, a gain $K$ is calculated from {\it
DD} robust stabilization with an external perturbation. The scenario
is composed as follows: 7 additive sources (UDP protocol) send 1000
bytes packet length with a 1Mbytes/s throughput between $t=40s$ and
$t=45s$. With the DD robust controller, the response is perturbed.
The closed-loop system converges to the same reference, the queue
size is stable and quickly regulated when the perturbation
disappeared.

\section{Conclusion}

In this preliminary work, we have proposed the construction of
robust AQMs for the congestion problem in communications networks.
The developed AQMs have been established by using Lyapunov theory
extended to delay systems and semi definite programming to solve the
Linear Matrix Inequalities. Note that the proposed methods have been
extended to the robust case where the delay in the loop is unknown. Finally, the AQMs have
been validated using NS simulator.

\bibliographystyle{plain}
\bibliography{mabiblio}
\end{document}